\documentclass[a4paper,11pt]{article}
\usepackage{etoolbox}
\usepackage{pos}
\newcommand{\MBS}{{M\"{o}bius Domain Wall Fermions}}

\newcommand{\hmu}{{\hat{\mu}}}
\renewcommand{\hookAfterAbstract}{%
  \par\bigskip
  \textsc{Preprint Number}: OU-HET-1253%
}
\title{Quark number susceptibility and conserved charge fluctuation for (2+1)-flavor QCD with Möbius domain wall fermions}

\ShortTitle{(2+1)-Flavor QCD with MDW Fermions}
\author*[a]{Jishnu Goswami}
\author[a]{Yasumichi Aoki}
\author[b]{Hidenori Fukaya}
\author[c,d]{Shoji Hashimoto}
\author[a]{Issaku Kanamori}
\author[c,d,e]{Takashi Kaneko}
\author[a]{Yoshifumi Nakamura}
\author[f]{Yu Zhang}

\onbehalf{(JLQCD collaboration)}

\affiliation[a]{%
 RIKEN Center for Computational Science (R-CCS),
Kobe 650-0047, Japan}

\affiliation[b]{%
Department of Physics, Osaka University, Toyonaka, Osaka 560-0043, Japan
}

\affiliation[c]{KEK Theory Center, High Energy Accelerator Research Organization (KEK), 
Tsukuba 305-0801, Japan}

\affiliation[d]{School of High Energy Accelerator Science, Graduate University for Advanced Studied (SOKENDAI),\\
Tsukuba 305-0801, Japan}

\affiliation[e]{Kobayashi-Maskawa Institute for the Origin of Particles and the Universe, Nagoya University,\\
 Aichi, 464-8603, Japan}

\affiliation[f]{Fakult\"at f\"ur Physik, Universit\"at Bielefeld, D-33615 Bielefeld,
Germany}

\emailAdd{jishnu.goswami@riken.jp}

\abstract{We present quark number susceptibilities and conserved charge fluctuations for (2+1)-flavor QCD using {\MBS} with a pion mass of \(135~\rm{MeV}\). Our results are compared with hadron resonance gas models below the QCD transition temperature and with \(\mathcal{O}(g^2)\) perturbation theory at high temperatures. Additionally, we compare our findings with results from staggered fermion discretizations. Furthermore, we also present results of leading order Kurtosis of electric charge and strangeness fluctuations.}

\FullConference{The 41st International Symposium on Lattice Field Theory (LATTICE2024)\\
 28 July - 3 August 2024\\
Liverpool, UK\\}

\begin{document}
\maketitle
\section{Introduction}

Exploring the phase diagram of quantum chromodynamics (QCD) at finite temperatures and densities is one of the fundamental problems in particle physics. The expected transition is from a hadronic phase, composed of hadronic bound states at low temperatures and densities, to a partonic phase of deconfined quarks and gluons at high temperatures and densities. At zero chemical potential, this transition is well established as a smooth crossover~\cite{Aoki:2006we}. Theoretical models suggest that at high baryon densities, the crossover could turn into a true phase transition, potentially featuring a critical point in the phase diagram~\cite{Stephanov:2004wx}.

Facilities such as the Relativistic Heavy Ion Collider (RHIC) and the Large Hadron Collider (LHC) realize the extreme conditions necessary to study QCD matter at finite temperatures and densities through heavy-ion collisions. These experiments probe the phase diagram, complementing theoretical studies using lattice QCD. In particular, lattice QCD simulations provide insights into the QCD phase structure through the analysis of quark number susceptibilities and fluctuations of conserved charges, which reflect the changing degrees of freedom near the transition.

This study investigates the QCD phase diagram using Möbius Domain Wall Fermions (MDWF)~\cite{Brower:2012vk, Kaneko:2013jla, Aoki:2021kbh, Aoki:2023fmn}, which ensure good control over chiral symmetry at finite lattice spacings. We present results for second-order fluctuations along the line of constant physics (LCP) corresponding to a continuum pion mass $135~\rm{MeV}$. These results are compared with non-interacting Hadron Resonance Gas models (PDGHRG and QMHRG2020) to explore the sensitivity of second-order observables to the hadron spectrum below the QCD pseudo-critical temperature. Furthermore, we compare our findings with results obtained using staggered fermions and $\mathcal{O}(g^2)$ perturbation theory. 
Finally, we construct leading-order expansion coefficients for the kurtosis of electric charge and strangeness fluctuations, which are relevant to locate the QCD critical point in the experimental data~\cite{Stephanov:2011pb}. In the following sections, all the dimensionful lattice parameters are expressed in lattice units, except otherwise stated explicitly.

\section{Calculation of quark number fluctuations up to fourth order with {\MBS}}\label{sec:qnsmeas}

Thermodynamic properties of QCD matter at finite temperature and density are often characterized by quark number fluctuations, which are derivatives of the pressure with respect to quark chemical potentials. For QCD with two light flavors (\(u, d\)) and one strange flavor (\(s\)), the pressure \(P\) can be expanded as a Taylor series in terms of the chemical potentials \(\mu_f\) (\(f = u, d, s\)):
\begin{align}
    \frac{P}{T^4} &= \frac{1}{VT^3} \ln Z(T, V, \vec{\mu}) = \sum_{i,j,k=0}^{\infty} \frac{\chi_{ijk}^{uds}}{i!j!k!} \hat{\mu}_u^i \hat{\mu}_d^j \hat{\mu}_s^k, \label{Pdefinition}
\end{align}
where \(T\) is the temperature, \(V\) is the spatial volume, and \(\hat{\mu}_f = \mu_f / T\) are the dimensionless chemical potentials. The coefficients \(\chi_{ijk}^{uds}\), known as generalized susceptibilities at vanishing chemical potentials, are derivatives of the logarithm of the partition function \(Z(T, V, \hat{\mu}_f)\) with respect to the chemical potentials:
\begin{align}
\chi_{ijk}^{uds} &= \frac{1}{VT^3} \left. \frac{\partial^{i+j+k} \ln Z(T, V, \hmu_f)}{\partial \hat{\mu}_u^i \, \partial \hat{\mu}_d^j \, \partial \hat{\mu}_s^k} \right|_{\hmu_f=0}, \quad \text{for } i + j + k \text{ even}; \label{chiTaylor} \\
\chi_{ijk}^{uds} &= 0, \quad \text{for } i + j + k \text{ odd}.
\end{align}

To implement the chemical potentials \(\hmu_f\) in the MDWF action, the temporal gauge links are modified as follows~\cite{Bloch:2007xi,Brower:2012vk}:
\[
(1 \pm \gamma_4) U_{\pm 4}(x) \rightarrow (1 \pm \gamma_4) e^{\pm \hmu_f} U_{\pm 4}(x),
\]
. The partition function can then be expressed as:
\begin{align}
    & Z = \int DU \prod_{f=u,d,s} \det \mathcal{M}(m_f, \hmu_f) \, \exp[-S_g], \nonumber \\
    & \det \mathcal{M}(m_f, \hmu_f)  = \frac{\det D(m_f, \hmu_f)}{\det D(m_{\text{PV}}, \hmu_f)}, \label{eq:partitionfunc}
\end{align}
where \(D(m_f, \hmu_f)\) is the MDMF Dirac operator for quark flavor \(f\) with mass \(m_f\) and chemical potential \(\hmu_f\), and \(m_{\text{PV}}\) is the Pauli-Villars regulator mass.

The quark number susceptibilities are expressed in terms of derivatives of \(\ln \det \mathcal{M}(m_f, \hmu_f)\) with respect to \(\hmu_f\), defined as:
\begin{align}
    D_n^f & \equiv \left. \frac{\partial^n}{\partial \hmu_f^n} \ln \det \mathcal{M}(m_f, \hmu_f) \right|_{\vec{\mu}=0}. \label{eq:Dn}
\end{align}
Using these definitions, the second- and fourth-order quark number susceptibilities can be written as:
\begin{align}
\chi_2^f &= \frac{N_\tau}{N_\sigma^3} K_2^f, \quad \chi_{11}^{fg} = \frac{N_\tau}{N_\sigma^3} K_{11}^{fg}, \quad (f \neq g), \\
\chi_4^f &= \frac{1}{N_\tau N_\sigma^3} \left[ K_4^f - 3 (K_2^f)^2 \right], \\
\chi_{31}^{fg} &= \frac{1}{N_\tau N_\sigma^3} \left[ K_{31}^{fg} - 3 K_2^f K_{11}^{fg} \right], \quad (f \neq g), \\
\chi_{22}^{fg} &= \frac{1}{N_\tau N_\sigma^3} \left[ K_{22}^{fg} - K_2^f K_2^g - 2 (K_{11}^{fg})^2 \right], \quad (f \neq g), \\
\chi_{211}^{fgh} &= \frac{1}{N_\tau N_\sigma^3} \left[ K_{211}^{fgh} - 2 K_{11}^{fg} K_{11}^{fh} - K_{11}^{gh} K_2^f \right], \quad (f \neq g \neq h).
\end{align}
Here, \(N_\tau\) and \(N_\sigma\) denote the temporal and spatial lattice sizes, respectively. The \(K\)-terms represent expectation values involving derivatives \(D_n^f\), as follows:
\begin{align}
K_2^f &= \langle (D_1^f)^2 \rangle + \langle D_2^f \rangle, \quad K_{11}^{fg} = \langle D_1^f D_1^g \rangle, \\
K_4^f &= \langle (D_1^f)^4 \rangle + 6 \langle (D_1^f)^2 D_2^f \rangle + 4 \langle D_1^f D_3^f \rangle + 3 \langle (D_2^f)^2 \rangle + \langle D_4^f \rangle. \\
K_{22}^{fg} &= \langle (D_1^f)^2 (D_1^g)^2 \rangle + \langle (D_1^f)^2 D_2^g \rangle + \langle  D_2^f (D_1^g)^2 \rangle + \langle D_2^f D_2^g \rangle, \\
K_{31}^{fg} &= \langle D_1^f (D_1^g)^3 \rangle + 3 \langle D_1^f D_2^f D_1^g \rangle + \langle D_3^f D_1^g \rangle, \\
K_{211}^{fgh} &= \langle (D_1^f)^2 D_1^g D_1^h \rangle + \langle D_2^f D_1^g D_1^h \rangle.
\end{align}

To relate the quark chemical potentials \((\mu_u, \mu_d, \mu_s)\) to conserved charges \((\mu_B, \mu_Q, \mu_S)\), we use:
\begin{align}
\mu_u &= \tfrac{1}{3} \mu_B + \tfrac{2}{3} \mu_Q, \quad \mu_d = \tfrac{1}{3} \mu_B - \tfrac{1}{3} \mu_Q, \quad \mu_s = \tfrac{1}{3} \mu_B - \tfrac{1}{3} \mu_Q - \mu_S. \label{eq:udsbqs}
\end{align}
The second-order conserved charge susceptibilities are then:
\begin{align}
\chi_2^B &= \frac{1}{9} \left( 2\chi_2^u + \chi_2^s + 2\chi_{11}^{ud} + 4\chi_{11}^{us} \right), \quad \chi_2^Q = \frac{1}{9} \left( 5\chi_2^u + \chi_2^s - 4\chi_{11}^{ud} - 2\chi_{11}^{us} \right), \quad \chi_2^S = \chi_2^s, \\
\chi_{11}^{BQ} &= \frac{1}{9}  \left(\chi_2^u - \chi_2^s + \chi_{11}^{ud} - \chi_{11}^{us} \right),
\quad \chi_{11}^{BS} = -\frac{1}{3}  \left(\chi_2^s + 2\chi_{11}^{us} \right),
\quad \chi_{11}^{QS} = \frac{1}{3}  \left(\chi_2^s - \chi_{11}^{us} \right).  
\end{align}

As shown in our earlier work~\cite{Goswami:2024kcq}, the term \((D_1^f)^2\) contributes significantly to noise in these susceptibilities. However, degeneracy in the \(u, d\) quarks (\(D_1^u D_1^d = (D_1^u)^2\)) partially cancels these contributions in \(\chi_2^Q\), making it less noisy than \(\chi_2^B\), \(\chi_{11}^{BQ}\), or \(\chi_{11}^{BS}\). Observables involving the strange quark (\(\chi_2^S, \chi_{11}^{QS}\)) are typically less noisy due to the heavier strange quark mass.

Finally, fourth-order fluctuations can also be transformed into conserved charge fluctuations using Eq.~(\ref{eq:udsbqs}).
\section{Lattice setup and parameters}
The gauge configurations for (2+1)-flavor QCD are generated employing MDWFs with $b=3/2$ and $c=1/2$~\cite{Brower:2012vk} for dynamical quarks.
The lattice size in the fifth dimension is $L_s=12$, and the domain-wall height is set to $M_5=1$ taking advantage of 3-level stout-link smearing applied for the gauge links.
This parameter choice is aimed at minimizing chiral symmetry violations.~\cite{Hashimoto:2014gta,Colquhoun:2022atw}

All simulations are conducted along a line of constant physics (LCP). Namely, the bare strange quark mass ($m_s$) is adjusted to keep the strange quark mass in physical units constant with changing temperatures. We chose the mass ratio of light to strange quark mass:
\(m_l/m_s = m_l^{\text{latt}}/m_s^{\text{latt}} = 1/27.4\), where $m_l$ and $m_s$ represent the bare light and strange quark masses, respectively, and $m_f^{\text{latt}}$ indicates the multiplicatively renormalizable lattice mass for each flavor $f\in\{l,s\}$. This ratio corresponds to a pion mass of $135~\mathrm{MeV}$ in the continuum. In particular, we use the relation, $m_s^{\text{phys}} = Z_m m_q^{\text{latt}} a^{-1}(\beta)$, where $m_s^{\text{phys}}=92~\mathrm{MeV}$ and $\beta$ is the gauge coupling. (The procedures for determination of $a(\beta)$ and $Z_m(\beta)$ are given in \cite{Aoki:2021kbh}, where the $\overline{\text{MS}}$ scheme at the renormalization scale $\mu=2~\mathrm{GeV}$ is used to determine $Z_m$.)

Due to the finite $L_s$, the input bare quark masses($m_f$) receive additive corrections, $m_f^{\text{latt}}=m_f+m_{\text{res}}$, where $m_{\text{res}}$ is the residual mass. We use the value of $m_{\text{res}}$ obtained in another series of simulations at the mass ratio, $m_l^{\text{latt}}/m_s^{\text{latt}}=1/10$, as $m_{\text{res}}$ is nearly independent of the light quark mass.

We have approximately $20,000$ trajectories for each temperature, and we perform measurements on every $100$ trajectories.

\section{Results and Discussion}
\subsection{Second-order quark number susceptibilities}

\begin{figure}[tbp]
    \centering
    \includegraphics[width=0.43\textwidth]{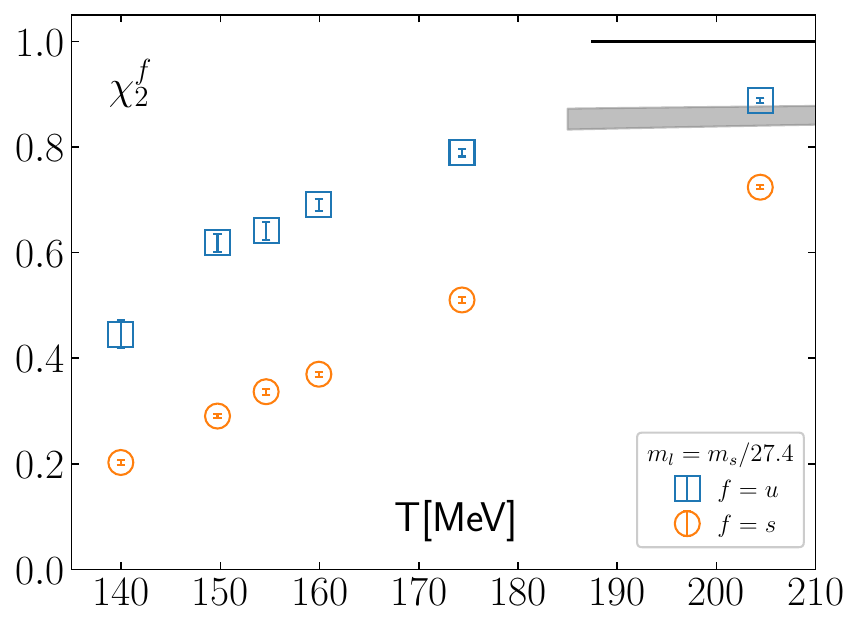}
    \hspace{0.02\textwidth}
    \includegraphics[width=0.43\textwidth]{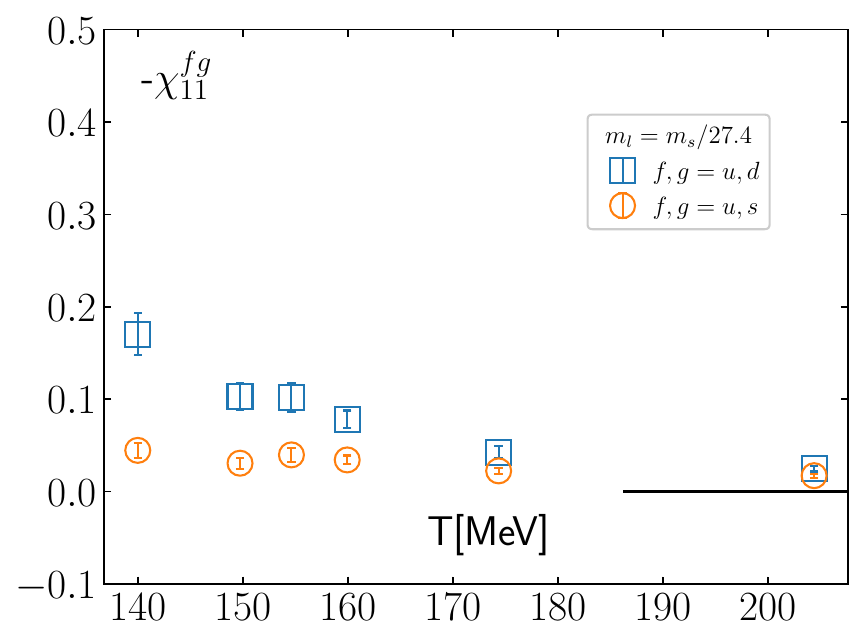}
    \caption{Diagonal (left) and off-diagonal (right) quark number susceptibilities for the coarse lattice ($N_{\tau}=12$) with a physical light quark mass ($m_l=m_s/27.4$) along the line of constant physics. The black line represents the free quark gas. The gray band is from $\mathcal{O}(g^2)$ perturbation theory.}
    \label{fig:qns_second_order}
\end{figure}

In Figure~\ref{fig:qns_second_order}, we present second-order quark number susceptibilities as a function of temperature, computed in (2+1)-flavor QCD for a light quark mass \(m_l = m_s/27.4\). The diagonal quark number susceptibilities increase smoothly as a function of temperature within the explored temperature range. In contrast, the off-diagonal susceptibilities stays relatively smaller compared to the diagonal susceptibilities. This increase may signals the deconfinement transition, where the degrees of freedom changes from hadrons to quark and gluons. The non-zero off diagonal quark number susceptibilities (flavor correlations) observed above \(T_{\mathrm{pc}}\) suggest interactions among quarks in the QGP phase.

We compare our results with the ideal gas limit at high temperatures, where QCD matter behaves as a weakly interacting gas of hadrons and gluons. The deviations from the ideal gas limit in diagonal quark number susceptibilities are well captured by \(\mathcal{O}(g^2)\) high-temperature perturbation theory~\cite{Vuorinen:2003fs, Mogliacci:2013mca,Bazavov:2013uja,Ding:2015fca,Bellwied:2015lba}. The off-diagonal quark number susceptibilities do not receive perturbative corrections at $\mathcal{O}(g^2)$ so they approach the free gas limit faster. In Figure~\ref{fig:qns_second_order}, we show both the ideal gas limit (black line) and \(\mathcal{O}(g^2)\) corrections (gray band), using a two-loop running coupling \(g^2(T)\) with a renormalization scale \(k_T \pi T\) for \(4 \le k_T \le 8\)~\cite{Bollweg:2022rps}. These results are consistent with previous staggered discretization calculations for physical quark masses~\cite{Borsanyi:2011sw,Bazavov:2011nk}.

At temperatures below \(T_{\mathrm{pc}}\), the degrees of freedom of QCD matter are ordinary hadrons, thus we discuss the results in terms of second-order conserved charge fluctuations. We compare with two hadron resonance gas (HRG) models: PDGHRG, which includes only 3- and 4-star resonances from the PDG booklet, and QMHRG2020~\cite{Bollweg:2021vqf}, which adds unobserved hadrons from quark models, along with 1- and 2-star PDG resonances.

\begin{figure*}[tbp]
\includegraphics[scale=0.30]{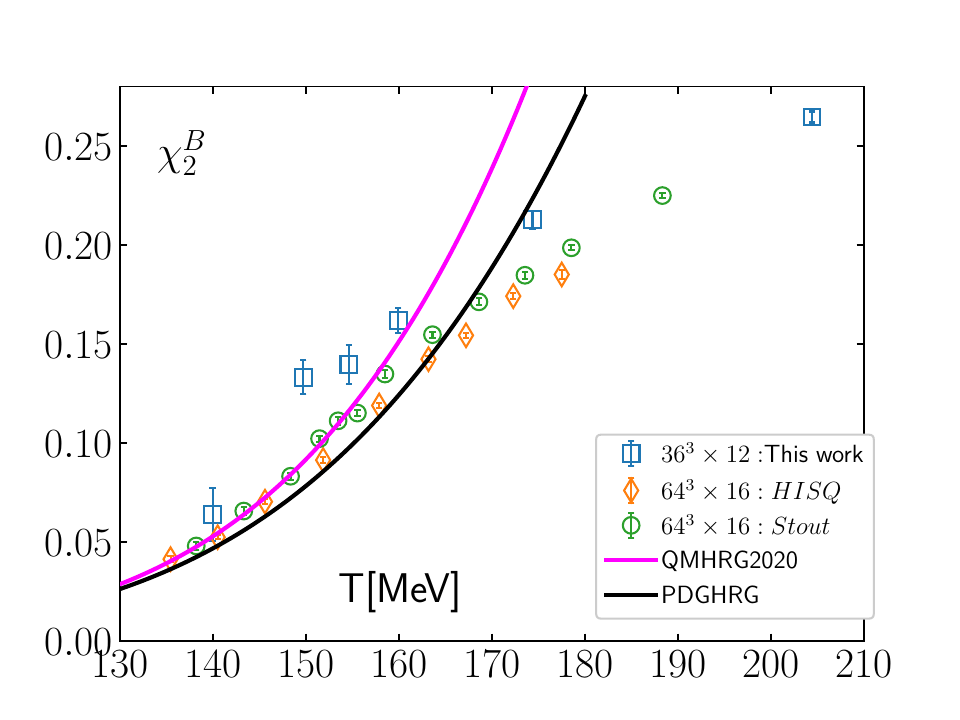}
\includegraphics[scale=0.30]{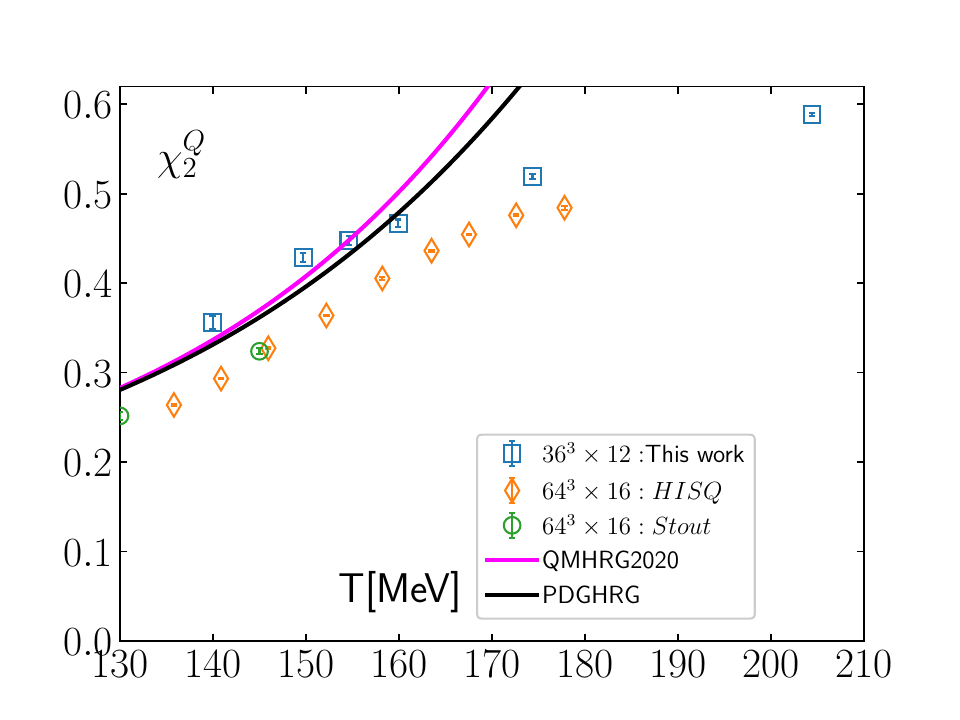}
\includegraphics[scale=0.30]{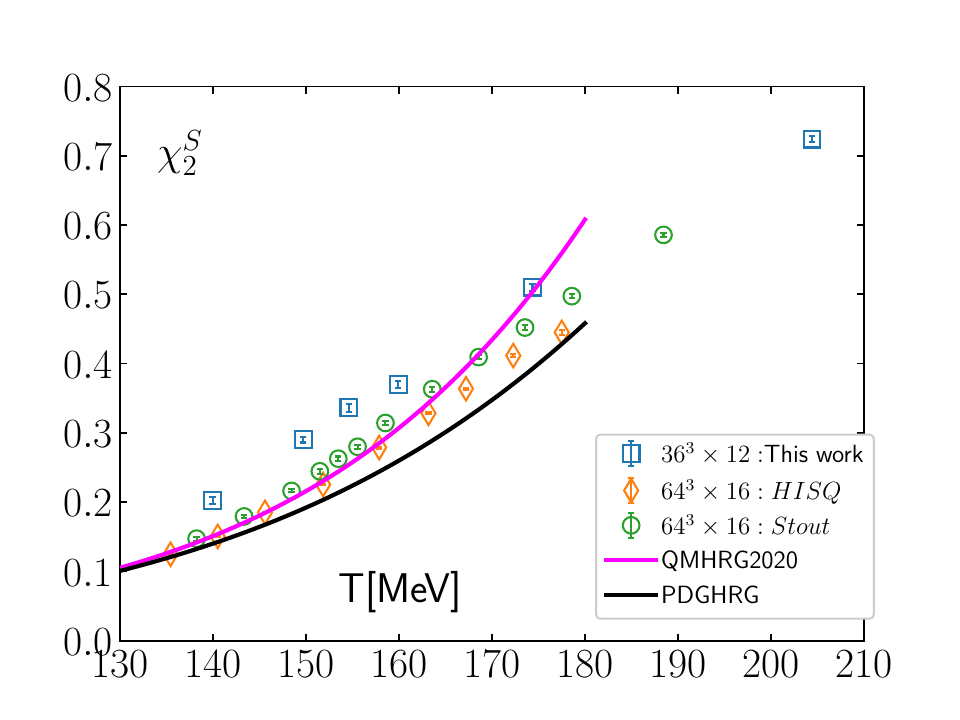}
\includegraphics[scale=0.30]{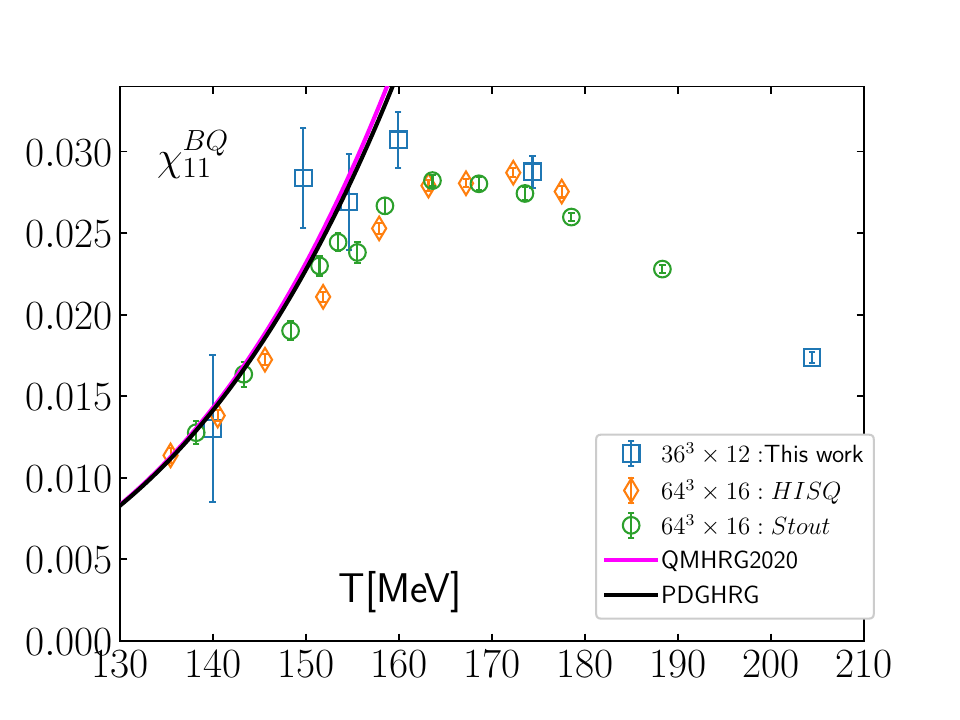}
\includegraphics[scale=0.30]{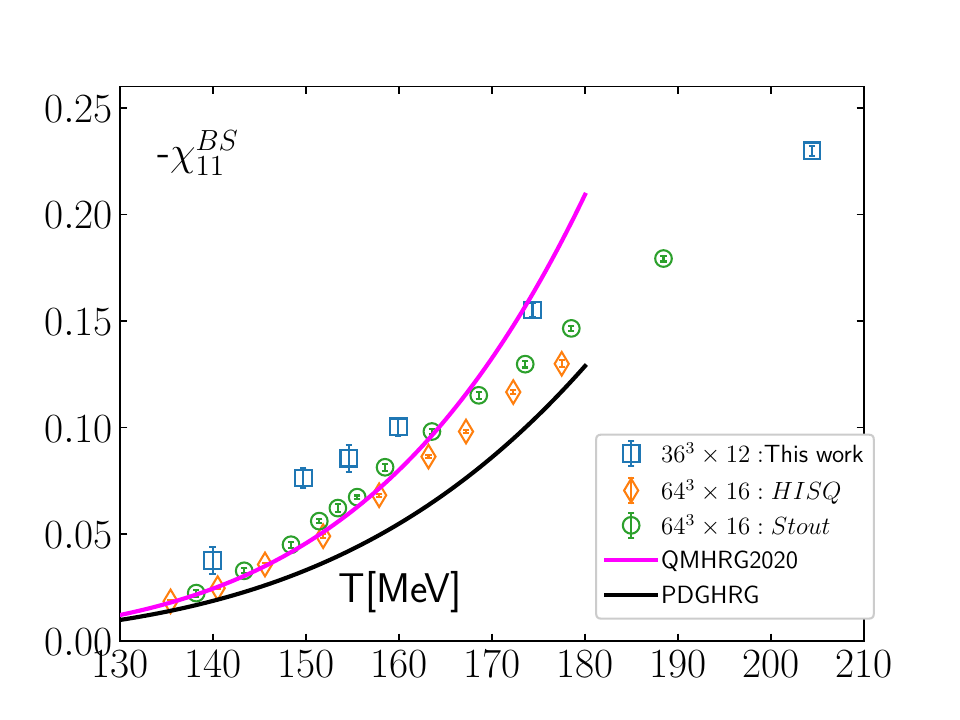}
\includegraphics[scale=0.30]{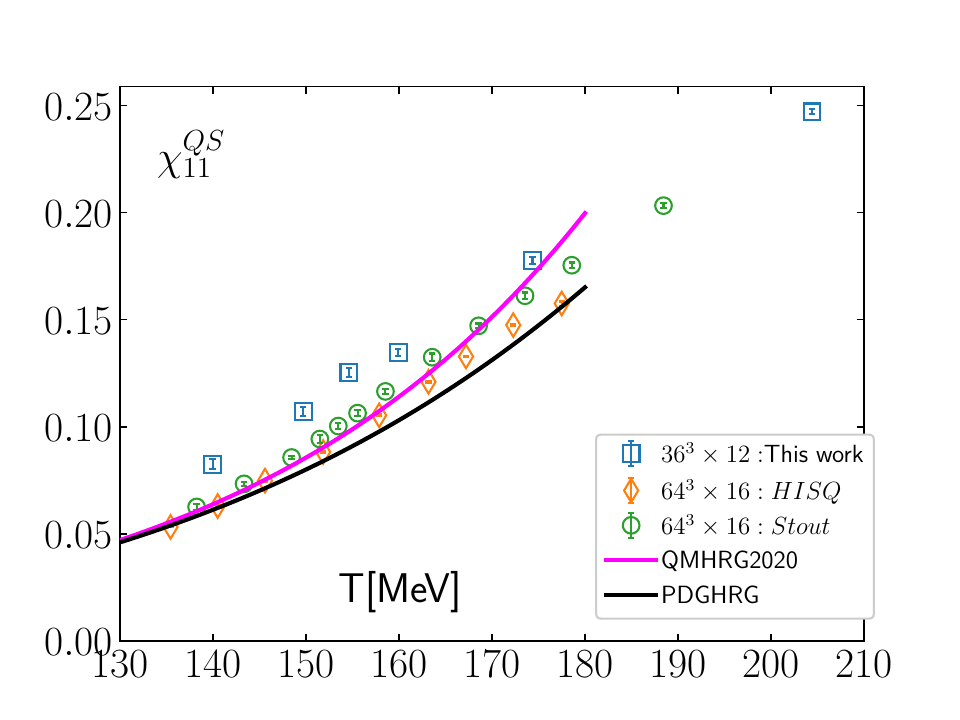}
\caption{Second order conserved charge cumulants 
 compared with staggered discretization scheme at finite lattice. We also compare the results with PDGHRG and QMHRG2020 model calculations. The HISQ data are taken from \cite{Bollweg:2021vqf} and the stout data are taken from \cite{Bellwied:2015rza}.
}
\label{fig:mlphycssecondeleccharge}
\end{figure*}

In Figure~\ref{fig:mlphycssecondeleccharge}, we compare second-order cumulants from MDWF calculations with those from staggered fermion calculations at finite lattice spacing, as well as with PDGHRG and QMHRG2020 predictions. In the strangeness sector, the inclusion of additional strange particles in QMHRG2020 improves agreement with lattice QCD. However, in the non-strange sector, the inclusion of extra particles does not result in a significant difference compared to PDGHRG.
For the electric charge susceptibility \(\chi_2^Q\), a discrepancy is observed between MDWF and staggered fermion results at temperatures below 160~MeV. Notably, MDWF results agree more closely with HRG models in this temperature range. In future, we will study these results with an additional lattice spacings.

\subsection{Fourth quark number susceptibilities and conserved charge fluctuations}
\begin{figure}[tbp]
    \centering
    \includegraphics[width=0.43\textwidth]{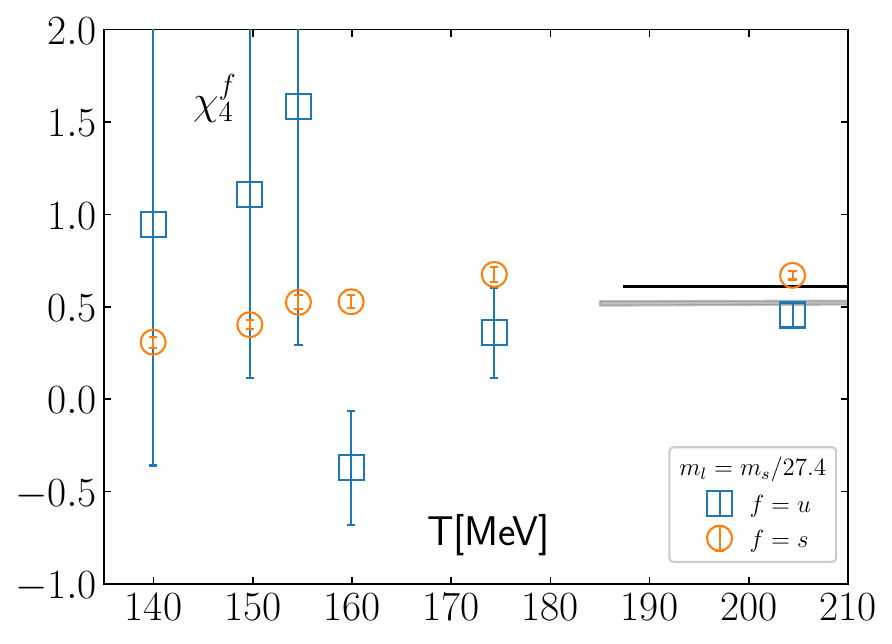}
    \hspace{0.02\textwidth}
    \includegraphics[width=0.43\textwidth]{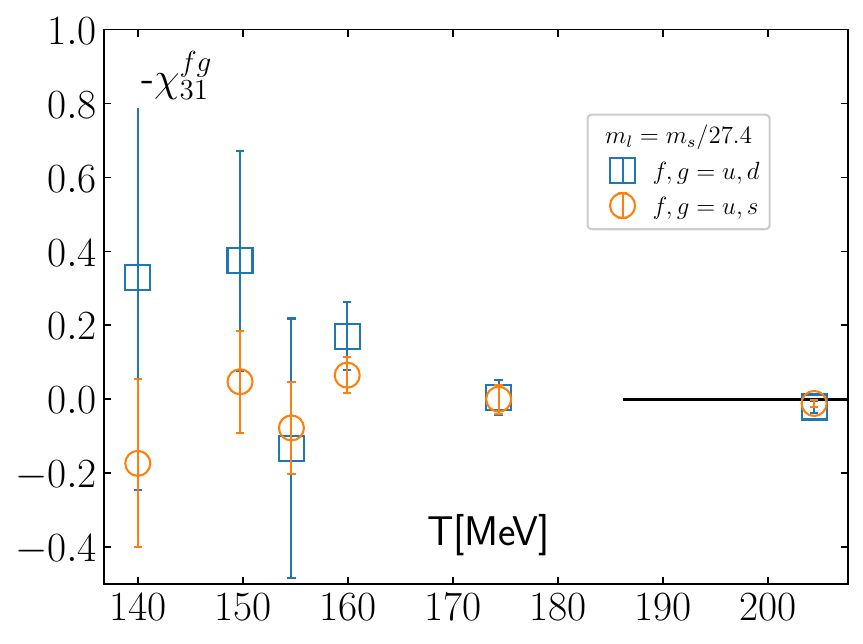}
    \caption{Diagonal (left) and off-diagonal (right) fourth order quark number susceptibilities for the coarse lattice ($N_{\tau}=12$) with a physical light quark mass ($m_l=m_s/27.4$) along the line of constant physics. The black line represents the free quark gas and. The gray band is from $\mathcal{O}(g^2)$ perturbation theory.}
    \label{fig:qns_fourth_order}
\end{figure}

In Figure~\ref{fig:qns_fourth_order}, we present the results of fourth-order quark number fluctuations. These are important quantities for constructing the fourth-order fluctuations which serves as NNLO Taylor expansion coefficient of the pressure. We acknowledge that calculating higher-order susceptibilities introduces substantial statistical noise, particularly in the hadronic phase, where fluctuations are suppressed. Despite using a large number of Gaussian random sources and employing dilution techniques, some observables, especially those involving light quark derivatives, exhibit larger uncertainties. We observe that the higher the derivative with respect to the light quark chemical potential (\(\mu\)), the noisier the data are. However, at high temperatures, the susceptibilities are well under control, and they approach the free quark gas limit and/or $\mathcal{O}(g^2)$ in perturbation theory.

\subsection{Ratio of fourth order to second order}\label{sec:cnfratiocomp}
The ratio of fourth-order to second-order cumulants of electric charge and strangeness, \(\chi_4^X/\chi_2^X\), could serve as an observable for determining the freeze-out parameters and/or searching for the QCD critical point. Here, we present the leading-order (LO) Taylor expansion, where the kurtosis of the electric charge ratio and strangeness ratio is expressed as:
\begin{equation}
R^{X}_{42}(\hat{\mu}_B, T) = \chi_4^X/\chi_2^X + O(\hat{\mu}_B^2), \quad X = Q, S.
\end{equation}

\begin{figure}[tbp]
    \centering
    \includegraphics[width=0.43\textwidth]{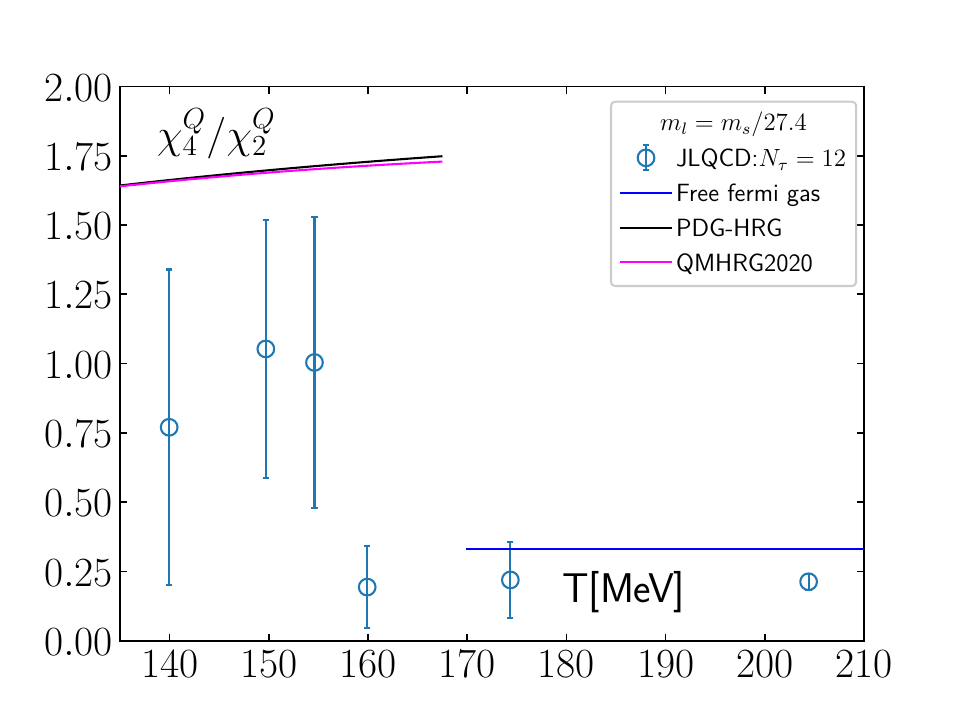}
     \hspace{0.02\textwidth}
    \includegraphics[width=0.43\textwidth]{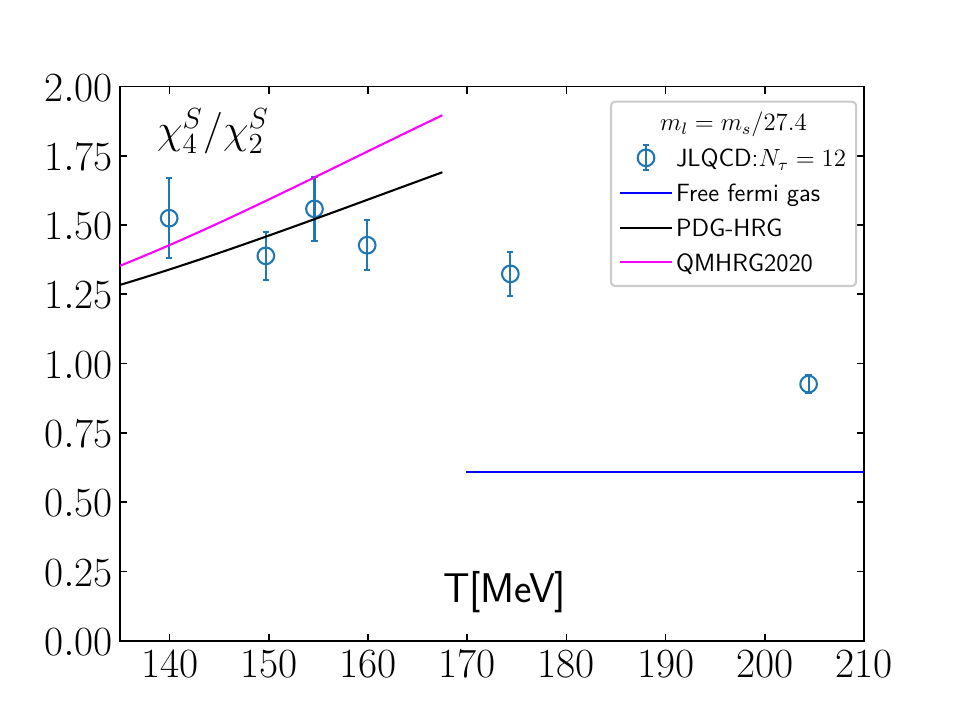}
    \caption{Leading order $R_{42}^X$, where $X=Q$ (left) and $S$ (right) calculated using MDWF.}
    \label{fig:kurtosis}
\end{figure}

We compare the leading-order kurtosis with predictions from the hadron resonance gas (HRG) models in Figure~\ref{fig:kurtosis}, specifically the PDGHRG and QMHRG2020. Our analysis shows that, within the current errors, \(R_{42}^S\) is consistent with both HRG models. However, the HRG model calculations of \(R_{42}^Q\) overshoot the lattice data.

Our preliminary results for \(R_{42}^X\) at \(\mu_B=0\), in the vicinity of the pseudo-critical temperature for (2+1)-flavor QCD, are:
\begin{eqnarray}
    R_{42}^Q = 
\begin{cases} 
1.05 \pm 0.46 & \text{for } T = 149.7~\text{MeV} \\
1.00 \pm 0.53 & \text{for } T = 154.6~\text{MeV}.
\end{cases} ~ ; ~
R_{42}^S = 
\begin{cases} 
1.38 \pm 0.09 & \text{for } T = 149.7~\text{MeV} \\
1.56 \pm 0.12 & \text{for } T = 154.6~\text{MeV}.
\end{cases}
\end{eqnarray}
Typically, experimental data from the STAR \cite{STAR:2014egu} and PHENIX \cite{PHENIX:2015tkx} collaborations for \(R_{42}^Q\) have large errors. Similarly, our lattice data also exhibits substantial errors, making it difficult to perform statistically significant comparisons at present. However, future experimental results with smaller errors could provide interesting comparisons.

\section{Summary and Outlook}
We presented our preliminary results of quark number susceptibilities and conserved charge fluctuations for (2+1)-flavor QCD using {\MBS}, comparing lattice results with HRG models and staggered fermion discretizations. At low temperatures, our results aligned with HRG predictions, consistent with hadronic degrees of freedom, while at high temperatures, they approached \(\mathcal{O}(g^2)\) high-temperature perturbation theory, indicating a transition to partonic matter. Fourth-order susceptibilities, although statistically noisier, were well controlled above \(T_{\mathrm{pc}}\) and matched the free theory limit. In particular, the \(\chi_2^Q\) results from {\MBS} were larger than those from staggered fermions at low temperatures but closer to the predictions of the HRG model. In the future, we will improve statistical precision for higher-order susceptibilities and include calculations at smaller lattice spacings for continuum extrapolations.

\section*{Acknowledgements}
The project is supported by the MEXT as ``Program for Promoting Researches on the Supercomputer Fugaku'' (JPMXP1020200105),``Simulation for basic science: approaching the quantum era"  (JPMXP1020230411) and Joint Institute for Computational Fundamental Science (JICFuS). I. K. acknowledges JPS KAKENHI (JP20K0396).
The simulations are performed on the supercomputer ``Fugaku'' at RIKEN Center for Computational Science (HPCI project hp240295,hp230207,hp200130, hp210165, hp220174 and Usability Research ra000001). We used the Grid~\cite{Boyle:2015tjk,githubGrid} for configuration generations and Bridge++~\cite{Ueda:2014rya,Aoyama:2023tyf, githubBridge} for measurements. We use AnalysisToolbox \cite{Altenkort:2023xxi,githubAnalysistoolbox} for statistical data analysis. J.G. would like to thank the long-term workshop on HHIQCD2024 at the Yukawa Institute for Theoretical Physics (YITP-T-24-02) for giving him a chance to deepen his understandings.
\bibliographystyle{JHEP}
\bibliography{bibliography}
\end{document}